\documentclass[fleqn,10pt]{wlscirep}
\usepackage[T1]{fontenc}
\title{Linking the connectome to action: Emergent dynamics in a robotic model of {\em C. elegans}}

\author[1,2,+]{Carlos E.~Valencia Urbina}
\author[3,+]{Sergio A.~ Cannas}
\author[1,4,+,*]{Pablo M. Gleiser}
\affil[1]{Medical Physics Department, Centro At\'omico Bariloche, R\'{\i}o Negro 8400, Argentina.}
\affil[2]{Instituto Balseiro, Universidad Nacional de Cuyo, R\'{\i}o Negro 8400, Argentina.}
\affil[3]{Instituto de F\'{\i}sica Enrique Gaviola (IFEG), Facultad de Matem\'atica, Astronom\'{\i}a, F\'{\i}sica y Computaci\'on, Universidad Nacional de C\'ordoba, Ciudad Universitaria, (5000), C\'ordoba, Argentina.}
\affil[4]{Universidad Nacional de R\'{\i}o Negro, R\'{\i}o Negro 8400, Argentina.}

\affil[*]{gleiser@cab.cnea.gov.ar}

\affil[+]{these authors contributed equally to this work}

\begin{abstract}
We analyse the neural dynamics and its relation with the emergent behaviour of a robotic vehicle that is controlled by a neural network numerical simulation based on the nervous system of the nematode {\em Caenorhabditis elegans}. The robot interacts with the environment through a sensor, that transmits the information to sensory neurons, while motor neurons outputs are connected to wheels. This is enough to allow robot movement in complex environments, avoiding collisions with obstacles.  Working with a robotic model makes it possible to keep track simultaneously of the detailed microscopic dynamics of all the neurons and also register the actions of the robot in the environment in real time. This allowed us   to study the interplay between connectome and complex behaviors. We found  that some basic features of the global neural dynamics and their correlation with behaviour observed in the worm appear spontaneously in the robot, suggesting they are just an emergent property of the connectome.
\end{abstract}
\begin{document}

\flushbottom
\maketitle

\thispagestyle{empty}

\section*{Introduction}

Understanding how the nervous system of living organisms encodes, organizes and sequences behaviours is one of the fundamental questions in biology and science in general \cite{branson2015imaging}.  The main challenge resides in the ability to monitor whole, or almost whole neural systems while registering activity \cite{anderson2014toward,datta2019computational}. With this goal in mind, many studies have  focused on model animals with small nervous systems, such as the fruit fly {\em Drosophila melanogaster} \cite{robie2017mapping}, the zebrafish {\em Danio rerio}  \cite{ahrens2013whole,kim2017pan,cong2017rapid,symvoulidis2017neubtracker} and the worm {\em Caenorhabditis elegans} ({\em C. elegans}) \cite{corsi2015transparent,kato2015global, kaplan2020nested}.  

The nematode {\em C. elegans} stands out in neuroscience studies as the first animal whose complete connectome has been mapped \cite{white1986structure}. Using serial electron microscopy  synapse-level neural maps have been constructed both for adult male and hermaphrodite \cite{white1986structure,cook2019whole}. By adding the sizes of the synaptic connections between cell pairs, and assuming that average synapse size is larger for stronger connections than for weak ones,  the map can be represented as an adjacency matrix with weights that quantify the amount of physical connectivity between pairs.  This allows for the description of the nervous system as a weighted graph.  

The abstraction of a neural systems into a set of nodes and weighted edges allows for the development of a theoretical framework to study the general organizing principles of the neural structures. This approach has proven to be successful \cite{sporns2004organization,sporns2010networks,fornito2016fundamentals,van2016comparative}, revealing many non trivial topological features, that  
are shared by nervous systems across species, such as network motifs \cite{sporns2004organization}, community structures \cite{sohn2011topological}, rich clubs \cite{towlson2013rich}, and small world network structure \cite{watts1998collective,varshney2011structural}. Also, it establishes the first step in the study of  the relation between the network structure and function, that is, on the  dynamical processes that can run on these structures. For example, the small world network structure is characterized by a high clustering and a short distance between nodes. This allows for the coexistence of a functional segregation in well defined regions while also allowing for a fast transfer of information consolidating  global integration into coherent states \cite{sporns2004organization}. 

The possibility of advancing beyond structural analysis, incorporating experimental information on neural dynamics has also been possible in {\em C. elegans}, thanks to calcium imaging techniques \cite{schrodel2013brain} that allow  recording  activity simultaneously in a large fraction of neurons. In a recent work, Kato {\em et al.} \cite{kato2015global} registered the activity of neurons in the head ganglia of worms while simultaneously recording their behaviour. They observed the presence of well defined clusters with synchronized activity. 
Also, using dimension reduction mathematical techniques they observed that the activity at brain level evolves into cycles on a low dimensional attractor-like manifold, where different segments, corresponding to the activities of different neuronal sub populations, can be mapped to represent the action sequences of the nematode \cite{kato2015global}. 
Kaplan {\em et al.} \cite{kaplan2020nested} extended this analysis to the entire nervous system, including head ganglia, ventral nerve cord and tail ganglia, finding similar results. Also, they found that the dynamics presents a hierarchical structure across brain and motor circuits, where slower dynamics constrain  the state and function of faster ones.  By registering the body postures and locomotor behaviour using automatic video tracking, they showed how the hierarchy  is used to coordinate behaviours across different time scales \cite{kaplan2020nested}.
 
These experimental results highlight the importance of studying  the nervous system, the body and the environment, as a coupled system 
in order to understand the properties that emerge from their continuous dynamical interaction \cite{chiel1997brain,webb2000does,webb2002robots,floreano2014robotics,clark1998being}. In this context the use of robots appears as an attractive modelling tool, since it is possible to access and have full control over the parameters and dynamical variables that govern their behaviour \cite{pfeifer2007self,izquierdo2013connecting}. Also, the physical implementation of a robot allows for testing the performance of algorithms in a body that is subject to the laws of physics and is immersed in real time in a natural environment. 

In this work we use a robot vehicle that is controlled by a neural network numerical simulation based on the {\em C. elegans} connectome
\cite{busbice2014extending}. This allows us to analyse the neural dynamics that emerges through the connectome,  and at the same time to study the interplay between the neural dynamics and the actions of the robot. 
We find that some basic features of the global neural dynamics of the worm, such as the presence of clusters of synchronized neurons \cite{kato2015global},  and  a nested hierarchical structure that couples slow and fast oscillating neurons \cite{kaplan2020nested},  can be explained just as an emergent consequence  of the connectome architecture, without need of any other modulatory mechanism.
Also, as observed in the worm \cite{kato2015global} the global neural activity in the robot evolves on a low-dimensional attractor like manifold.  When the trajectories in the manifold are contrasted with the behaviour of the robot, we also observe cyclical dynamics that represents action sequences.   The interplay between the emergent dynamics and the actions of the robot highlights the key role that the connectome plays in the most basic aspects of locomotor activity in {\em C. elegans}.

\section*{Results}

For the experiments we use robots based on the design originally presented by T. Busbice \cite{busbice2014extending}.
The robots are vehicles with one distance sensor and two wheels,  which allow them to sense the environment and move on the ground.  
The microscopic dynamical units of the model, the neurons, evolve in a numerical simulation with discrete time steps. At each time step all the neurons add their input signals up to a given threshold value. When a neuron  surpasses  the threshold  it fires, distributing the signal to its neighbours, and resetting its state to zero. 

The neural network controlling the robot is based on the {\em C. elegans} connectome.  The neurons in the nervous system of the worm  can be divided into three categories according to their neuronal structural and functional properties: sensory neurons, interneurons and motor neurons \cite{white1986structure,varshney2011structural}.  In the robot, the distance sensor activates a number of sensory neurons when the distance from the robot to an obstacle is below a given threshold. The wheels are controlled by the output of motor neurons.  With this setup we conducted experiments where the robot was allowed to roam freely in a room with random obstacles, recording  simultaneously the individual dynamics of all the neurons and also the actions of the robot

We chose this particular design for our experiments due to its attractive features. On the one hand, the design of the robot allows for a straightforward construction that  can be easily reproduced with low cost. On the other hand, the neural simulation that controls the robot has elementary dynamical units, and uses the biological information of the  connectome for their interaction. With this stripped down approach T. Busbice showed that the robot presents emergent behaviours, which allow it to spontaneously navigate and avoid obstacles  \cite{busbice2014extending,busbiceYoutube}.  Thus, the robotic model constitutes an excellent platform to study the interplay between the connectome and emergent behaviours. However, up to now neither the dynamics of the model, nor its correlation with behaviour has been analysed. The main objective of the present work is to fill that gap, as well as to contrast, when possible, the results obtained in the robot with those coming from recent experiments on {\em C. elegans}.

\subsection*{Emergent neural dynamics}
\subsubsection*{Frequency and phase synchronization}

\begin{figure}
\includegraphics[scale=0.5]{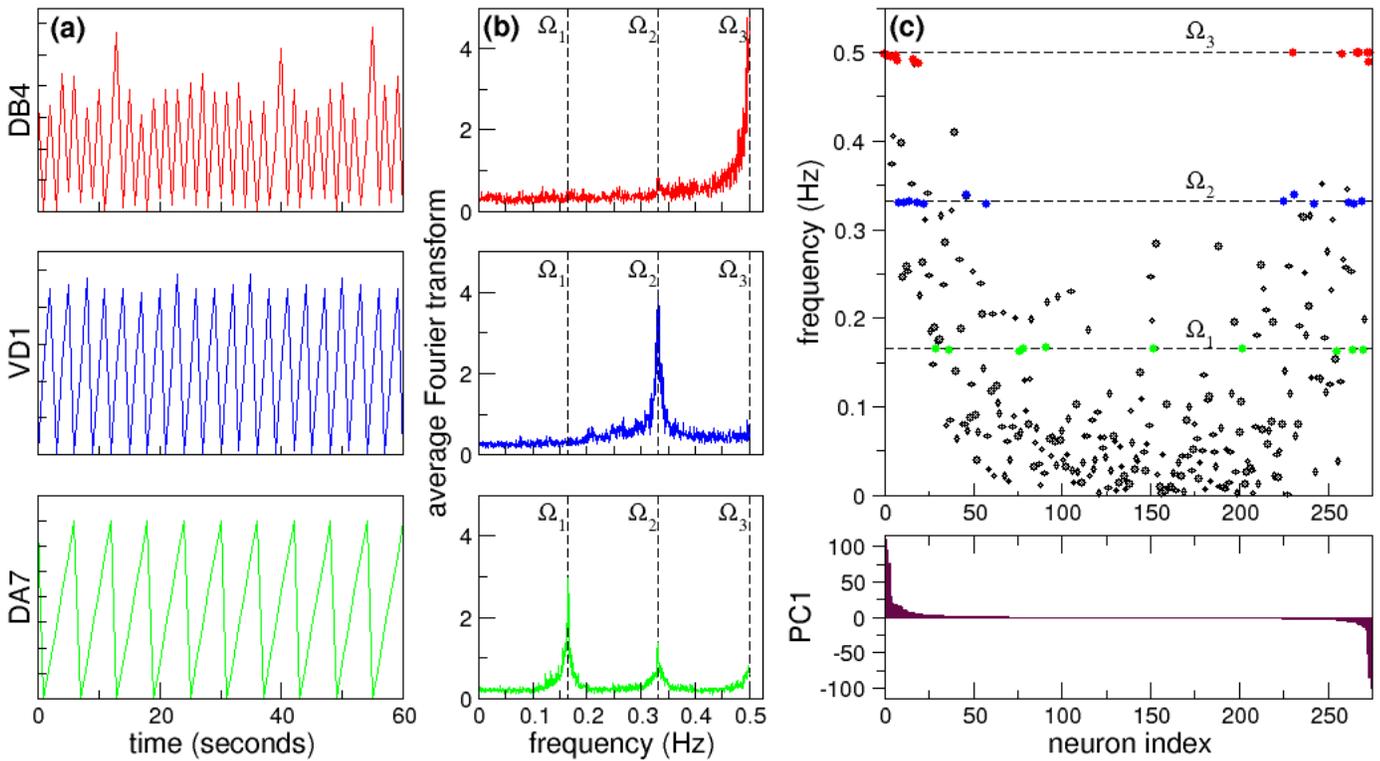}
 \caption{\label{Fig1} {\bf Neuronal dynamics and frequency synchronization}. (a) Dynamics of three ventral cord motor neurons, DB4 (top, red), VD1 (middle, blue),  and DA7 (bottom, green). The figure shows the neural signals as a function of time in the same $60$ second time interval.  (b) Average Fourier transforms of neurons in three synchronized clusters for a full $15$ minute long experiment. (c)  Characteristic frequency of all the neurons sorted according to the corresponding PC1 weight (bottom). The three synchronized clusters are highlighted using the same colors as in (b). }
 \end{figure}

It is worth stressing that in the robotic model all the neurons have the same firing threshold, and thus have identical individual dynamics as isolated units. However, we expect their dynamics to reflect the non-uniform distribution of synaptic connectivity. Neurons in a central position,  with a large number of inputs or receiving connections with large weights  can reach the threshold faster, and thus fire frequently. On the other hand, nodes in a peripheral position will take longer to reach the threshold, and thus have a slower dynamics.
In fact, we observed that as soon as the robot is turned on and begins to explore the environment, the neurons present oscillating dynamics with different frequencies. As an example we plot in Fig. \ref{Fig1} (a) the signals of three ventral cord motor neurons in a fixed time interval.  We  quantitatively characterize the oscillations by performing a  Fourier transform (FT). We find that the FT present sharp peaks (see Methods and Supplementary Fig. S1). We define the characteristic frequency  of the neurons, $\omega$, as the highest peak in the FT.

When we extend our focus from individual to collective dynamics we observe that some neurons are clustered in groups that share the 
same characteristic frequency. Even more, in some cases the complete shape of their FT, including smaller peaks,  overlap. This allows us to define synchronized clusters as groups of neurons with overlapping FT.  In Fig. \ref{Fig1} (b) we plot the average Fourier Transform of neurons in three different frequency synchronized clusters. The averaged Fourier transforms have sharp peaks, that allows us to define the clusters $\Omega$ by their characteristic frequency:  $\Omega_1=0.165 \pm 0.002$ Hz (green, bottom), $\Omega_2=0.332 \pm 0.003$ Hz (blue, middle) and  $\Omega_3 = 0.496 \pm 0.004$ Hz (red, top)  (see also Supplementary Tables S1A-S1C  and Figures S2A-S2C for a complete list of neurons in each cluster, their individual signals and corresponding FTs). 
 
In {\em C. elegans} clusters of neurons that present coordinated dynamics have been registered experimentally by Kato {\em et al.} \cite{kato2015global}. Using calcium imaging techniques they  recorded neural activity with single cell resolution in all the head ganglia and some ventral cord motor neurons. In these experiments approximately 100 neurons were scanned three times per second in 20 minute long runs, generating high dimensional datasets. Using principal component analysis \cite{jolliffe2016principal} for dimensional reduction they were able to cluster neurons with correlated signals.  In particular, they produced neuron weight vectors (PCs), showing that clusters with opposite signs in their PCs oscillate in antiphase, and according to their weight vector sign are correlated with either forward or backward behaviour of the worm. 
 
With this idea in mind we also performed  a principal component analysis, producing neuron weight vectors (PCs) for the neural time series of all the neurons in the robot. This allowed us to advance further in the quantitative characterization of the frequency synchronized clusters. In Fig. \ref{Fig1} (c) we plot the characteristic frequencies of all the neurons, highlighting  the three frequency synchronized clusters  $\Omega_1$ (green),  $\Omega_2$  (blue), and $\Omega_3$ (red) already presented in Fig. \ref{Fig1} (b).  The neurons have been sorted according to their  first principal component weight (PC1) (Fig.~\ref{Fig1} (c), bottom),  i.e., the leftmost neurons have the highest positive value, while the rightmost have the lowest negative PC1.   With this sorting the neurons with highest frequencies are in the extremes, while the lower frequencies are in between with small  PC values.

Note that all the synchronized clusters involve neurons including both positive and negative PC1 values. The cluster with lower frequency, $\Omega_1$, presents a broad distribution of neuron weights. For higher frequencies the neurons present a stronger segregation towards extreme PC1 values. In fact, both $\Omega_2$ and $\Omega_3$  are clearly divided into two smaller subclusters: one with only positive and another with only negative PC1 values.  This segregation reflects differences in the firing times of the neurons. In each cluster the firing times of the neurons are proximal. In contrast, when one compares the signals of neurons between clusters a shift is observed.  The largest shift occurs for extreme PC values, when the neurons in the different subclusters of the same frequency oscillate mostly in antiphase (see Supplementary Fig. S3). 
 
\subsubsection*{Nested neuronal dynamics}

We analyse now how the collective oscillations in the  frequency synchronized clusters are also coupled between themselves. In Fig. \ref{Fig1} (b) we showed the average Fourier transforms of neurons in three frequency synchronized clusters. Note that the average FT of the cluster with the lowest frequency, $\Omega_1$, also has two smaller peaks,  that coincide with the characteristic frequency of the other clusters: $\Omega_2$ and $\Omega_3$. As a guide to the eye we indicate with dashed lines these frequencies in the three panels. This reveals a coupling between the different frequency synchronized clusters. 

\begin{figure}
\includegraphics[scale=0.35]{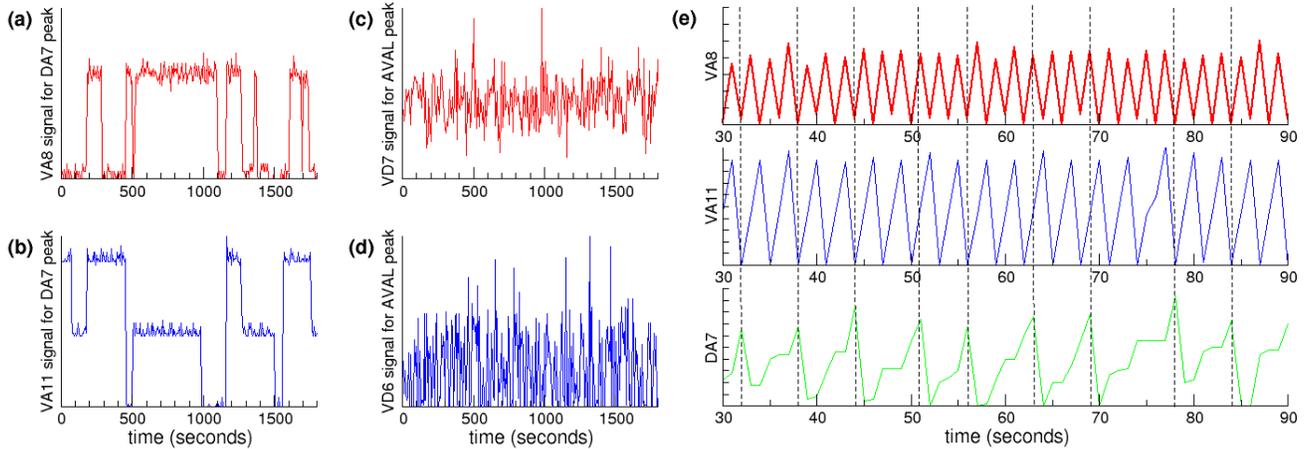}
 \caption{\label{Fig2} {\bf Nested neuronal dynamics}. (a) VA8 (red) signal  and  (b) VA11 (blue) when DA7 reaches a maximum. (c) VD7 (red)  and (d) VD6 (blue)  when AVAL reaches a maximum.(e) The signals of VA8, VA11 and DA7 in a $60$ second time interval. Vertical dashed lines are a guide to the eye to show the hierarchical relation.}
 \end{figure}

In order to quantitatively study the coupling between oscillations at different frequencies, we analysed the signals of neurons in a given cluster  when the signal of a neuron in another cluster with a lower characteristic frequency reaches its maximum \cite{jensen2007cross}. If there is a coupling between the oscillations, then we expect that the signal of the neuron with highest frequency could present small fluctuations around the same mean value every time the maximum with lowest frequency is reached. On the contrary, if there is no coupling, then the signal is expected to vary randomly. 
In Fig. \ref{Fig2} we  plot the signals of neurons in  $\Omega_2$ and $\Omega_3$ when a  neuron selected  from the cluster with lower characteristic frequency $\Omega_1$ reaches its maximum. 
In particular, we focused our attention in the neurons with  the highest positive PC1 weight in each of the synchronized clusters. Figs. \ref{Fig2} (a) and (b) show that these neurons are coupled, as the same mean value persists in extended time intervals. Also, to visualize this result  we plot in Fig. \ref{Fig2} (e) the signals of these neurons in a fixed time interval, using vertical dashed lines as a guide to the eye. The figure clearly shows a nested hierarchical relation between oscillations at different frequencies.  
In Figs. \ref{Fig2} (c) and (d)  we plot the signals of the neurons that have the lowest negative PC1 weights. In sharp contrast to neurons with positive weights, the signals are not correlated, and fluctuate throughout the whole experiment.

Kaplan {\em et al.} \cite{kaplan2020nested} observed a nested neuronal dynamics in experiments that extended the  work of Kato {\em et al.} \cite{kato2015global}  from the head of {\em C. elegans}  to the entire nervous  system, including the ventral nerve cord and tail ganglia.  They showed that this dynamical structure allows for the coordination of behaviours at different time scales. Moreover, they showed that the nested dynamics persists when the animals are immobilized and cannot execute the behaviours, concluding that  it is an intrinsic property that emerges from the neurons and their circuit interactions. The results obtained with the robot support that conclusion. 

\subsection*{Neural dynamics and robot actions}

\begin{figure}
\includegraphics[scale=0.6]{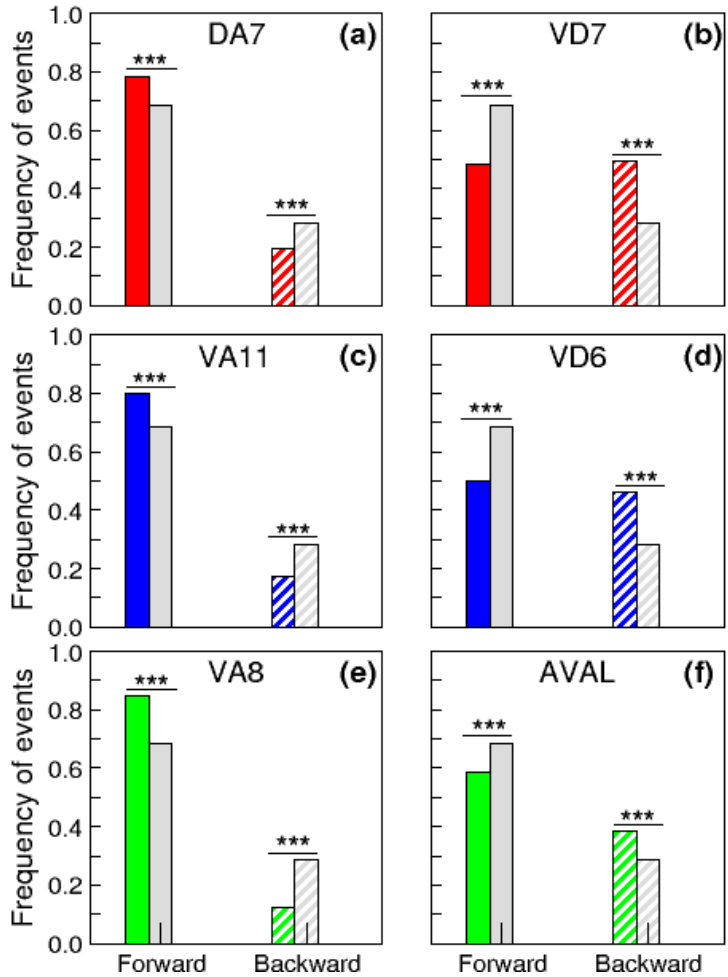}
 \caption{\label{Fig3} {\bf Correlation of neural dynamics with robot actions}. Fraction of forward (continuous bars) and backward (dashed bars) events registered when 
the selected neurons fired.  (a) DA7, (c) VA11, and (e) VA8 correspond to the neurons with the largest positive PC1 in the synchronized clusters $\Omega_1$, $\Omega_2$ and $\Omega_3$ , while (b) VD7, (d) VD6, and (f) AVAL correspond to the neurons with the lowest negative PC1 in these clusters. The gray bars correspond to the fraction of forward and backward events in a $20$ minute long experiment. A chi-square test ($p-value  < 0.00001$, significant at $p < .01$) showed that the opposing effects of all these neurons in the actions of the robot are significant.}
 \end{figure}

In this section we analyse the correlation between the emergent neural dynamics and the actions of the robot. The robotic model is well suited for this study, since it  allows for recording of neural  dynamics while simultaneously registering  actions. To determine if there was a correlation between a given action and the activity of specific neurons, we contrasted the time series of the neurons with the actions the robot was executing. First, we registered all the action events in a 20 minute long experiment. Then, we analysed which actions the robot was executing when a given neuron had fired, that is, when its value was reset to zero. Finally, we contrasted the fractions of events in this two time series to see if there were significant variations. In the experiments the robot moves mostly forward and backward, while the turnings are usually short events where the robot only changes its direction, so we focused mainly on forward and backward events. 
We found that the neurons  in the synchronized clusters with the largest positive PC1 weights promote forward events, while the neurons with the lowest negative weights promote backward events.  Fig. \ref{Fig3} shows the fraction of events registered when the neurons with the largest positive and lowest negative PC1 weights in the synchronized clusters fired (see also Supplementary Table S2). The coloured bars correspond to: $\Omega_1$ (green),  $\Omega_2$ (blue), and $\Omega_3$ (red). In each figure, the results are contrasted with the fractions of events for the whole experiment (grey bars). Note that a significant increase in forward events is observed when the neurons with the largest positive PC1 fire (First columns in Figs. \ref{Fig3} (a), (c) and (e)), while a reduction is observed in the neurons with the lowest negative PC1 (First columns in Figs. \ref{Fig3} (b), (d) and (f)). At the same time, an increase in backward events is observed when neurons with negative PCs fire (dashed columns in FIgs. \ref{Fig3} (b), (d) and (f)), while a decrease in the fraction of backward events is observed in neurons with positive weights (dashed columns in FIgs. \ref{Fig3} (a), (c) and (e)).  These results reveals the role that the segregation of the neurons in the  synchronized clusters plays in promoting different actions to be executed by the robot. Interestingly, in {\em C. elegans},  Kato {\em et al.} \cite{kato2015global} noted that neurons that promote opposing behaviours, such as backward and forward crawling, also have opposing signs of their PC1 weights. 

Now, we extend our attention from individual neurons to global dynamics and it relation with action. As in Ref. \cite{kato2015global} we build temporal PC time series  by taking weighted averages of each principal component with the full  multi-neural time series. This allows for a drastic dimensional reduction, where the temporal PCs have  signals that are representative of neuronal clusters.  

\begin{figure}
\includegraphics[scale=0.5]{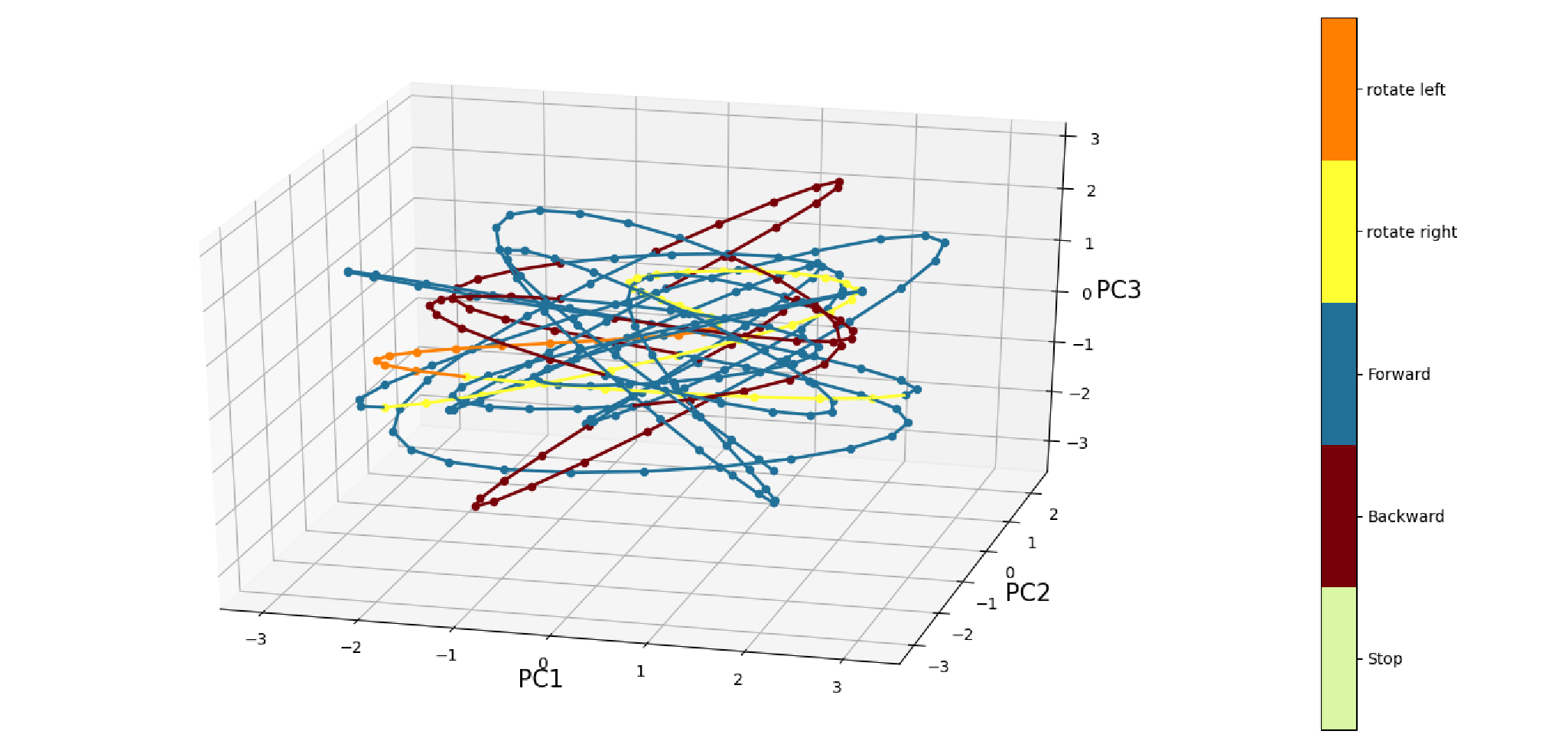}
 \caption{\label{Figure4} {\bf Global neural dynamics and actions}. Phase plots of the first three temporal PCs in a selected time interval. The actions of the robot are represented using different colors. The figure shows that the global dynamics evolves on a low dimensional attractor like orbit, where different segments of the trajectory correspond to well defined motor states.}
 \end{figure}

In Fig. \ref{Figure4} we show a parametric plot of  the time integrals of the first three temporal PCs in a fixed time interval. The trajectory presents a behaviour similar to a dynamical system in an attractor like orbit. Furthermore, we found that the trajectories for the complete time series of the experiments always presented this behaviour, and remained bounded in a global state characterized by the presence of cyclical orbits.  When the activity of the robot was taken into account, we found that different segments in the trajectories correspond to different actions. This can be clearly seen in Fig. \ref{Figure4}, where we used different colors to represent the action the robot was executing at a given time. We chose a particular time segment where the robot meets an obstacle, stops, reverses and turns \cite{busbiceYoutube}. Note that the different actions can be clearly distinguished as well defined segments of the trajectory. 

In {\em C. elegans}, Kato {\em et al.} \cite{kato2015global} found that the neural activity evolves on a low-dimensional attractor-like manifold. Also, different segments in this manifold, that correspond to the synchronization of different groups of neurons, are clearly correlated to different motor states.  Our results highlight the role that the connectome plays in the emergence of these attractor-like cyclical states. 

\section*{Discussion}

The nematode {\em C. elegans} is  a model organism that allows for an integrated view  where the continuous interaction between the nervous system, the body and the environment can be studied as a coupled dynamical system. In two seminal works Kato {\em et al.} \cite{kato2015global} and Kaplan {\em et al.} \cite{kaplan2020nested}  showed that organization of behaviour in the worm is encoded in a hierarchical structure of globally distributed, continuous, and low-dimensional neural dynamics. 
These results led to the conclusion that the behavioural states are encoded in the brain as an internal representation that emerges from the neurons and their circuit interactions. Our analysis of the neural dynamics and its correlation with the actions in a robotic model based on the connectome of the nematode {\em C. elegans} allowed us to test this hypothesis, at least as a first approximation. At this point it is worth stressing that our goal was not to establish a one-to-one comparison between the robotic vehicle and the worm. In fact, we chose to build robots following the idea proposed by T. Busbice due to its stripped down design. With this approach we have shown that  the interaction of  extremely  simple dynamical units through the complex network defined by the connectome allows for the emergence of a global dynamics that  presents a number of characteristics which are also observed in the worm. This include the emergence of synchronized clusters that can be correlated with the actions of the robot. It also includes a global state that evolves on an attractor-like trajectory, where different segments correspond to the actions the robot is executing. 

Surprisingly, we also observed the emergence of a nested hierarchical structure,  even when it could not be exploited by the robot. In the worm, Kaplan {\em et al.} \cite{kaplan2020nested} showed  that the presence of a  hierarchical structure in the neural dynamics of the worm allows for the coordination of behaviours across different time scales. This include head movements, body undulations, and also forward and reverse bouts. The design of the robot we used does not allow for the execution of some of this behaviours. Kaplan {\em et al.} \cite{kaplan2020nested} observed that the hierarchical structure persisted even when the animals were immobilized, concluding that it is an intrinsic property that emerges from the neurons and their circuit interactions. The results obtained with the robot highlight this conclusion, i. e.  that the presence of a nested dynamics is an emergent property of the interactions of the neurons through the connectome. 

Summarizing, we have shown that a number of characteristics observed in the neural dynamics  of {\em C. elegans} can be attributed to the complex network structure of the connectome. We expect that some of our conclusions can be extended to other connectomes, as a number of common principles have been identified in  the comparison of the topological layout of nervous systems across species \cite{van2016comparative}. At the same time, differences in the results  can shed light in understanding the effects of variations that are species specific. As larger neural networks and ever increasing detail on their dynamics are registered, the complexity and high dimensionality of the data will set new challenges for analysis and interpretation. For example the Drosophila hemibrain connectome involves approximately 25000 neurons, including regions involved in functions such as associative learning, fly navigation and sleep \cite{xu2020connectome}. Simple robotic models as the one we analysed here can provide a useful tool to test behavioural hypothesis relating neural structure and function. 

\section*{Methods}

In our experiments we used two different robotic vehicles with a similar design: A commercially available GoPiGo robot \cite{Dexter}, and a custom made robotic vehicle. Essentially the robots consists of a vehicle with two lateral motors connected to wheels and a distance sensor in the front. This allow the vehicles to move forward, backward and turn in any direction. We used 3V DC electric motors connected to $65mm$ plastic wheels with rubber tires.  The GoPiGo robot uses a GoPiGo 2 board for motor control, while the custom made uses a  L9110S dual motor driver.  Both robots use the  GoPiGO laser Distance Sensor, that is based on the VL53L0X chip. The sensor allows for very accurate distance measures, with a range up to two meters with a millimetre resolution. 

The control of the robot was implemented in a custom numerical simulation using the  Python 3 programming language, and is based on the original program developed by Timothy Busbice \cite{busbice2014extending} and the version implemented in the GoPiGo robot \cite{Github}.  The numerical simulation was run in a Raspberry Pi \cite{RaspberryPi} computer, that also interfaced with the distance sensor and the motor control boards. The program allows the neural signals and the actions of the robot to be sampled once per second,  setting the maximum frequency of the signals to be $\omega=0.5 Hz$. We used rechargeable battery power that allowed the robots tu run autonomously for up to one hour.

\bibliography{referencias_scirep}

\section*{Acknowledgements}

P.M.G. wishes to thank C. Buffa, M. G. Cascallares, and E. Scerbo, and M. Zimmer. S.A.C. acknowledges financial support by  CONICET (Argentina) through grants
PIP 11220150100285   and  SeCyT (Universidad Nacional de C\'ordoba, Argentina). P.M.G. acknowledges financial support by Agencia Nacional de Promoci\'on Cient\'{\i}fica y T\'ecnica through grant PICT 2016-1042. 

\section*{Author contributions statement}

C.E.V.U. and P.M.G. conceived and conducted the experiments, C.E.V.U., S. A. C. and P.M.G. analysed the results.  All authors reviewed the manuscript. 

\section*{Additional information}

The authors declare no competing interests.

\end{document}